\documentclass[12pt]{article}
\usepackage{amsmath,epsf,amssymb,latexsym,cite}
\usepackage[ps,dvips,matrix,arrow,frame,import,curve,color]{xy}

\newif\iffigs\figstrue

\DeclareFontFamily{U}{rsf}{}
\DeclareFontShape{U}{rsf}{m}{n}{
  <5> <6> rsfs5 <7> <8> <9> rsfs7 <10-> rsfs10}{}
\DeclareMathAlphabet\Scr{U}{rsf}{m}{n}

\def\O{\Scr{O}}
\def\C{{\mathbb C}}
\def\P{{\mathbb P}}

\def\Z{{\mathbb Z}}

\def\Hom{\operatorname{Hom}}

\def\sHom{\operatorname{\Scr{H}\!\!\textit{om}}}
\def\Ext{\operatorname{Ext}}
\def\End{\operatorname{End}}

\def\Tr{\operatorname{Tr}}

\def\dP#1{\mathrm{dP}_{#1}}
\def\cmod#1{\hbox{$#1$--\bf mod}}

\def\id{{\mathbf{1}}}

\def\cM{{\Scr M}}

\def\cA{{\Scr A}}
\def\cB{{\Scr B}}

\def\cD{{\Scr D}}

\def\cE{{\Scr E}}
\def\cF{{\Scr F}}

\def\cC{{\Scr C}}
\def\DC{\mathbf{D}}

\def\ff#1#2{{\textstyle\frac{#1}{#2}}}

\def\mf#1{\mathfrak{#1}}

\def\ddd{\mathsf{d}}

\begin{document}

\begin{titlepage}
\begin{flushright}
DUKE-CGTP-05-05\\
SLAC-PUB-11261\\
SU-ITP-05/21\\
hep-th/0506041\\
June 2005
\end{flushright}
\vspace{.5cm}
\begin{center}
\baselineskip=16pt
{\fontfamily{ptm}\selectfont\bfseries\huge
Superpotentials for Quiver Gauge Theories}\\[20mm]
{\bf\large  Paul S.~Aspinwall $^{1, 2, 3}$ and Lukasz M.~Fidkowski$^1$
 } \\[7mm]

{\small

$^1$ Department of Physics, Stanford University,
Stanford, CA 94305-4060\\ \vspace{6pt}
$^2$ SLAC,  Stanford, CA 94305-4060 \\ \vspace{6pt}
$^3$ Center for Geometry and Theoretical Physics, 
  Box 90318 \\ Duke University, 
 Durham, NC 27708-0318
 }

\end{center}

\begin{center}
{\bf Abstract} 
\end{center}
We compute superpotentials for quiver gauge theories arising from
marginal D-Brane decay on collapsed del Pezzo cycles $S$ in a
Calabi-Yau $X$. This is done using the machinery of $A_\infty$
products in the derived category of coherent sheaves of $X$, which in
turn is related to the derived category of $S$ and quiver path
algebras. We confirm that the superpotential is what one might have
guessed from analyzing the moduli space, i.e., it is linear in the
fields corresponding to the $\Ext^2$s of the quiver and that each such
$\Ext^2$ multiplies a polynomial in $\Ext^1$s equal to precisely the
relation represented by the $\Ext^2$.
\vspace{2mm} \vfill \hrule width 3.cm
\vspace{1mm}
%{\footnotesize \noindent email:
%psa@cgtp.duke.edu, \,  @stanford.edu}
\end{titlepage}

\vfil\break

%%%%%%%%%%%%%%%%%%%%%%%%%%%%%%%%%%%%%%%%%%%%%%%%%%%%%%%%%%%%%%%%

\section{Introduction}    \label{s:intro}

\def\fd#1{\mathsf{#1}}

Singularities of string backgrounds have attracted much attention and
have been investigated using a variety of methods
\cite{DM:qiv,DDG:wrap,DFR:orbifold,Douglas:Dlect,me:TASI-D,Franco:2005rj}.
One is to study the gauge theory on a $D$-brane probe of the
singularity.  While there has been much work done on extracting gauge
theory data for various types of singularities (abelian \cite{DM:qiv}
and non-abelian \cite{Johnson:1996py} orbifolds, conifolds
\cite{KW:coni,MP:AdS}, toric \cite{FHH:toric, Feng:2001xr, FHHU:tor-Sei,
Franco:2005rj} and generalized del Pezzo \cite{Feng:2002fv, Wijn:dP}
singularities), a general method for determining the superpotential
has been lacking.  In \cite{FHH:toric} the superpotential was obtained
from integrating the quiver relations for certain cases, with various
ad-hoc methods being used to resolve ambiguities that arise in such an
integration.  In this paper, using previous work of \cite{AK:ainf}, we
present a general rigorous method for obtaining the superpotential of
such quiver gauge theories from the quiver relations.  We show that
the superpotential is just the naive sum of terms of the form relation
times the $\Ext^2$ field corresponding to the relation.  We apply the
method to the trivial example of a $\P^2$ as well as to a $\dP 1$, in
which case we get a non-homogeneous superpotential.  In principle we
can apply it to a general del Pezzo singularity --- all we need is the
del Pezzo quiver and relations.

We deal with flat compactifications $M \times X$ where $M$ is $4$
dimensional Minkowski space and $X$ is a Calabi-Yau manifold, and we
probe the theory with space filling branes --- i.e., 
$(n+3)$-branes, where $n$ is the dimensionality of the brane
within $X$.  We expect such branes to be BPS and stable when probing a
smooth point of $X$, but to marginally decay into a collection of
so-called fractional branes when the point becomes singular.  We
consider singularities obtained when a complex surface (i.e., one that
has 4 real dimensions) $S$ shrinks down to zero size in $X$ by varying
the K\"ahler parameters.  Assuming that $S$ is smooth and irreducible,
it is known that $S$ must be a del Pezzo surface, i.e., $\P^1 \times
\P^1$, or $\P^2$ blown up at $m$ points (denoted by $\dP m$), where
$m$ ranges from $0$ to $8$.  The $3+1$ dimensional quiver gauge theory
associated to this marginal decay into fractional branes is the one
whose superpotential we are after.  This set up has been studied
extensively in the literature
\cite{AM:delP,MP:AdS,BGLP:dPo,BP:toric,FHHI:quiv,HI:quiv,CFIKV:,
Wijn:dP,HW:dib,Herz:exc}.

The moduli space of a D-brane is given by the space of critical points
of the superpotential. Thus, knowing the moduli space one may make a
guess at the form of the superpotential. In the case at hand, this leads
to a natural conjecture for the superpotential. By using more rigorous
methods we are able to show that this conjecture is correct.

For our analysis we will use the algebraic machinery of the derived
category of coherent sheaves, developed in particular in
\cite{Doug:DC,AL:DC,me:TASI-D}.  The fractional branes have tractable
representations as elements of the derived category $\DC(X)$ of
coherent sheaves on $X$.  In principle the method of
\cite{AK:ainf} can be applied to find the so-called
$A_\infty$ products in the algebra of $\Ext$ groups.  These $A_\infty$
products are determined by combinatorial relations that they have to
satisfy (coming from Feynman diagrams in the associated topological
theory) and they encode the superpotential.  Applying the technique of
\cite{AK:ainf} directly is difficult and in order to make the
problem tractable, we instead proceed in two steps.  First, we use a
spectral sequence argument to reduce the problem from one of studying
sheaves on $X$ to the simpler one of studying sheaves on $S$.  It is
in this reduction that we show that each $\Ext^2$ appears linearly in
the superpotential, multiplying a term that involves only $\Ext^1$s
and is determined by $A_\infty$ relations over $S$.  To compute these
we exploit the well understood properties of $\DC(S)$, and
specifically its intimate relation with the derived category of quiver
representations \cite{Bon:dPq, Bon:quiv}.  We see that the terms
involving the $\Ext^1$s are just the (possibly non-homogeneous)
relations in the quiver.  It is important to note that we obtain the
superpotential only up to certain nonlinear field redefinitions (see
\cite{AK:ainf}) --- this is the most that could be expected
from such topological sigma model methods as we use.

The plan of the paper is as follows: in section \ref{q:quivers} we
review quivers and sheaves on del Pezzo surfaces, and how they relate.
In section \ref{s:superp} we review $A_\infty$ algebras and the method
of \cite{AK:ainf} for using the topological $B$ model to compute
$A_\infty$ structure and hence the superpotential.  In section
\ref{sdP:sdP} we introduce the quiver gauge theory we want to study
and prove that its superpotential is linear in the $\Ext^2$s, which
multiply terms determined by the $A_\infty$ structure over $S$ ---
this is the reduction from sheaves on $X$ to sheaves on $S$.  In
section \ref{quiv:quiv} we solve the problem on $S$ by reframing it as
a computation in the derived category of quiver representations, and
apply the solution to the case $S=\P^2$ and the more nontrivial case
$S=\dP 1$.  This example illustrates the general algorithm that can be
carried through for any quiver with known relations.

\section{Quivers and Sheaves} \label{q:quivers}

\subsection{Quivers}

We now review the necessary mathematical notions relating to quivers
and their representations.  Further background can be found in
\cite{me:TASI-D}.  First, a quiver is a directed graph $Q$ that
consists of nodes $v_i$ and arrows $a_\alpha$.  Its path algebra $A$
is defined as follows: as a vector space, $A$ is generated by all of
the paths constructed through concatenation of arrows in $Q$.  The
product structure of $A$ is defined on these generators as follows: if
path $1$ ends on the same node that path $2$ begins on, the product is
defined to be the obvious concatenation; otherwise it is defined to be
$0$.  Note that corresponding to each node $v_i$ we have a
corresponding zero length path and hence an idempotent element $e_i$
of $A$.

In the remainder of the paper, we will deal with a slight
generalization to a quiver with relations.  This is just a quiver
whose path algebra is defined as the above $A$ quotiented out by a
subspace generated by linear combinations of paths called relations.
We stipulate that any given relation must be a linear combination of
paths between the same two nodes.  It does not, however, need to be
homogeneous.  A simple example (with homogeneous relations) is the
so-called Beilinson quiver, defined as:

\begin{equation}
\begin{xy} <1.0mm,0mm>:
  (0,0)*{\circ}="a",(20,0)*{\circ}="b",(40,0)*{\circ}="c",
  (0,-4)*{v_0},(20,-4)*{v_1},(40,-4)*{v_2}
  \ar@{<-}@/^3mm/|{a_0} "a";"b"
  \ar@{<-}|{a_1} "a";"b"
  \ar@{<-}@/_3mm/|{a_2} "a";"b"
  \ar@{<-}@/^3mm/|{b_0} "b";"c"
  \ar@{<-}|{b_1} "b";"c"
  \ar@{<-}@/_3mm/|{b_2} "b";"c"
\end{xy}  
\end{equation}with relations $a_\alpha b_\beta - a_\beta b_\alpha$.

For a given quiver $Q$, we can consider the associated category $\cmod
A$ of left $A$-modules.  For any left $A$-module $V$ we can form the
vector spaces $V_i = e_i V$, of dimension $N_i$; if we then think of
each $V_i$ as living on node $v_i$ then we see that multiplication by
any arrow $a_\alpha$ acts as a linear transformation between the
spaces $V_i$ at the tail and head of $a_\alpha$, and these linear
transformations respect the relations.  This structure is known as a
representation of the quiver $Q$, of dimension $(N_i)$.

A map $\phi$ between left $A$-modules $V$ and $W$ is simply a linear
transformation that commutes with the action of $A$, i.e., $\phi (av) =
a \phi (v)$.  If we think of $V$ and $W$ as quiver representations
then this condition is just the obvious constraint that the maps from
$V_i$ to $W_i$ must commute with the linear transformations induced by
the arrows, i.e., $\phi$ is a map of representations.  Thus we
sometimes refer to the category of left $A$-modules as the category of
quiver representations, and we use these terms interchangeably from
now on.

Corresponding to each node of $Q$ there are two distinguished
representations $P_i$ and $L_i$; when we make the connection to
sheaves on del Pezzo surfaces below, these will correspond to sheaves
in the exceptional collection and fractional branes, respectively, as
we will see.  $L_i$ is defined simply as the one dimensional
representation with $V_i = \C$ and all other $V_j = 0$.  $P_i$ is
defined as the subspace of $A$ generated by all paths that begin at
$v_i$; it is trivially seen to be a subrepresentation.  It may seem
that $P_i$ is a rather large representation, and indeed, if the quiver
has any loops there will be infinite dimensional $P_i$.  However, in
the case of quivers associated to del Pezzo's there will be an ordering
on the nodes that is respected by the arrows and hence all $P_i$ will
be finite-dimensional.  This finite dimensionality is in large part
responsible for the tractability of the problem and motivates the
reduction mentioned earlier from sheaves on the Calabi-Yau to sheaves
on the del Pezzo.  In the Beilinson quiver, for example, we see that
$P_0$, $P_1$, and $P_2$ have dimensions $(1,0,0)$, $(3,1,0)$, and
$(6,3,1)$ respectively.

When we make the connection between quivers and sheaves it will be
through the derived category.  Before we talk about that, however, let
us first discuss some basic homological properties of quivers.  First
of all, one can show that the $P_i$ are projective objects in $\cmod
A$.  In fact, they form a complete set in the sense that any left
$A$-module has a resolution by various direct sums of these $P_i$.
These projective resolutions can be used to compute higher $\Ext$
groups.  For example, the projective resolution of any $L_i$ is

\begin{equation}
\xymatrix@1{
\ldots\ar[r]&\bigoplus_{k}P_k^{\oplus r_{ik}}
  \ar[r]&\bigoplus_{k}P_k^{\oplus n_{ik}}\ar[r]&P_i\ar[r]&L_i\ar[r]&0.
}  \label{eq:Lres}
\end{equation}Here $n_{ij}$ is the number of arrows from node $i$ to
node $j$ and $r_{ij}$ is the number of independent relations imposed
on paths from $i$ to $j$.  In the case of the Beilinson quiver, the
resolutions are:

\begin{equation}
\xymatrix@R=3mm{  
&&0\ar[r]&P_0\ar[r]&L_0\ar[r]&0\\
&0\ar[r]&P_0^{\oplus3}\ar[r]&P_1\ar[r]&L_1\ar[r]&0\\
0\ar[r]&P_0^{\oplus3}\ar[r]&P_1^{\oplus3}\ar[r]&P_2\ar[r]&L_2\ar[r]&0\\
}
\end{equation}
Noting that $\Ext^k$ is the $k$th derived functor of $\Hom$ and that
$\Hom(P_i,L_j) = \delta_{ij}\C$ we can compute $\Ext^p(A,B)$ by taking
a projective resolution

\begin{equation}
\xymatrix@1{
\ldots\ar[r]&\Pi_2\ar[r]&\Pi_1\ar[r]&\Pi_0\ar[r]&A\ar[r]&0,
}  \label{eq:Pres}
\end{equation}
where the $\Pi_i$ are direct sums of $P_i$'s and from it constructing
the complex

\begin{equation}
\xymatrix@1{
0\ar[r]&\Hom(\Pi_0,B)\ar[r]&\Hom(\Pi_1,B)\ar[r]&\Hom(\Pi_2,B)
\ar[r]&\ldots
}
\end{equation}
The cohomology of this complex in the $p$th position is then
$\Ext^p(A,B)$.  Using this method one can show that

\begin{equation}
\begin{split}
  \dim\Ext^1(L_i,L_j) &= n_{ij}\\
  \dim\Ext^2(L_i,L_j) &= r_{ij}.
\end{split}
\end{equation}
In general, higher $\Ext$'s may also exist. For example, $\Ext^3$
represents relations amongst relations. However, for the purposes of
quiver gauge theories, the appearance of higher $\Ext$'s is unphysical
\cite{AM:delP,Herz:ouch} and so we assume
\begin{equation}
  \Ext^k(L_i,L_j) = 0,\quad k\geq3.
\end{equation}

Finally, we note that we can recover the quiver path algebra from the
projective representations $P_i$.  Supposing $Q$ is a quiver with $n$
nodes, this is done as follows: we let

\begin{equation}{T = P_0 \oplus P_1 \oplus \cdots \oplus P_{n-1}.}
\end{equation}
Using the fact that $\Hom(P_i,P_j)$ is simply the vector space of
paths from $j$ to $i$ we can verify that

\begin{equation}
  A \cong \End(T)^{\textrm{op}}.
\end{equation}
In other words, $A$ is just the algebra $\End(T)$ with the product
structure reversed.

\subsection {The Derived Category}

Having reviewed this preliminary material about quiver representations
we move on to briefly discuss the derived category.  As mentioned
above, the derived category will form a bridge between the quiver
representations that we have already discussed and the category of
coherent sheaves introduced below.  One source is \cite{me:TASI-D};
here we just review the facts we will use.

Given any abelian category $\cA$ (such as that of quiver
representations, or that of coherent sheaves discussed below) we can
define its derived category $\DC(\cA)$ as follows.  The objects in
$\DC(\cA)$ are complexes of objects in $\cA$:

\begin{equation}
\xymatrix@1{
  \ldots\ar[r]&\cE^0\ar[r]^{d_0}&\cE^1\ar[r]^{d_1}&
      \cE^2\ar[r]^{d_2}&\ldots,
}
\end{equation}
To construct the morphisms in $\DC(\cA)$, we begin with the abelian
group of all possible maps between complexes (not necessarily
respecting the differential).  These maps are graded by their degree
$p$ and can be written as

\begin{equation}
{\sum_{n} f_{n,n+p}}
\end{equation}
where $f_{m,n}$ is a map from $\cE^m$ to $\cE^n$.  We define a
differential on this group by (abusing notation slightly):

\begin{equation}
{(df)}_{n,p+1} = d_{n+p} f_{n,p} - (-1)^p f_{n+1,p} d_n
\end{equation}
The derived morphisms are now defined as the cohomology of this group,
with formal inverses added in for all quasi-isomorphisms (that is,
those chain maps which induce isomorphisms on cohomology).

Now we state some necessary results without proof.  Given any object
$A$ in $\cA$, we can construct the associated one term complex whose
only nonzero entry is $A$, at the zeroth position.  For brevity we
will henceforth refer to both the object and the associated one term
complex by $A$.  Then, for $A, B$ in $\cA$, $\Ext^{\bullet}(A,B)$ is
given by the group of derived morphisms between the complexes
associated to $A$ and $B$, with the grading on $\Ext$ corresponding to
the grading of the derived morphisms.  In fact, this is the
generalization of the notion of $\Ext$ to the arbitrary elements of
$\DC (\cA)$.  Also, any $A$ in $\cA$ is equivalent to its projective
or injective resolution in the derived category.  Further, if we
represent either $A$ by its projective resolution or $B$ by its
injective resolution then the generators of $\Ext(A,B)$ can be written
as honest chain maps between these complexes.

\subsection {Sheaves}  \label{ss:shvs}

We now turn to reviewing key aspects of the other relevant category,
that of coherent sheaves.  It turns out that (as we will see in more
detail below) the derived category of coherent sheaves on a Calabi-Yau
manifold $X$, denoted $\DC(X)$, precisely describes $D$-branes in the
topological B model defined on $X$.  The open string modes stretching
between them are described by the $\Ext$ groups of the sheaf homs
between the relevant branes \cite{Doug:DC,AL:DC,me:TASI-D}.  These, in
turn, describe the massless spectrum of the physical theory on $M
\times X$.  In fact $\DC(X)$ contains enough information to determine
the tree-level superpotential of the low energy effective theory, in
the form of $A_\infty$ products.  We will discuss all of this below,
but for now let us start by introducing coherent sheaves on
Calabi-Yau's and del Pezzo's.

The category of coherent sheaves on a space $X$ is an enlargement of
the category of vector bundles (also referred to as ``locally free
sheaves'') on $X$ --- it contains vector bundles as well as all
kernels and cokernels of maps of vector bundles.  For a precise
definition, starting from the general concept of a sheaf, see
\cite{me:TASI-D} or \cite{Hartshorne:}.  We can very roughly think of
it as including, in addition to vector bundles over $X$, more exotic
objects such as vector bundles over submanifolds of $X$.

In the physical problem we consider $D$-branes on a shrinking cycle
$S$ which is embedded in $X$: $i: S \rightarrow X$.  $i$ induces an
embedding $i_*: \DC(S) \rightarrow \DC(X)$, and it is no surprise that
the branes we'll be interested in are in fact in the image of $i_*$.
Now, $\DC(S)$ has been studied extensively by mathematicians and is
well understood.

We proceed by first defining a complete strongly exceptional
collection of sheaves on $S$ to be an ordered set $\{\cF_0, \cdots,
\cF_{n-1}\}$ that generates $\DC(S)$ and satisfies $\Ext^p_S (\cF_i,
\cF_j) = 0$ for $p \neq 0$ and any $i$ and $j$, and $\Ext^0_S(\cF_i,
\cF_j) = \Hom_S (\cF_i, \cF_j) = 0$ for $i>j$ and $\Hom_S(\cF_i,
\cF_i) = \C$.  Given such a complete strongly exceptional collection,
we can define

\begin{equation}
{A = \End (\cF_0 \oplus \cF_1 \oplus \cdots \oplus
  \cF_{n-1})^{\textrm{op}}} 
\end{equation}
It turns out that $A$ is the path algebra of a quiver $Q$, and the
$\cF_i$ are isomorphic (as $A$-modules) to the projective
representations $P_i$ defined above.  Given this we can reconstruct
the quiver uniquely simply by noting that $\Hom_S (\cF_i, \cF_j) =
\Hom (P_i, P_j)$ is just the space of paths from node $j$ to node $i$.
In fact, Bondal \cite{Bon:dPq} proved that the derived category of
$A$-modules, $\DC(\cmod A)$ is equivalent to $\DC(S)$.

As an example, consider $S = \P^2$.  An exceptional collection is
given by $\{\O, \O(1), \O(2)\}$.  We have $\Hom(\O, \O(1)) \cong
\C^3$, $\Hom(\O(1), \O(2)) \cong \C^3$.  Denote these maps, which are
just multiplication by the homogeneous coordinates, by $x_i$ and $y_i$
respectively, $i = 1,2,3$.  We also have $\Hom(\O, \O(2)) \cong \C^6$
--- these maps are multiplication by homogeneous degree two polynomials
in the homogeneous coordinates.  All this implies that we have three
arrows $x_i$ from node $2$ to node $1$, three arrows $y_i$ from node
$3$ to node $2$, and that all paths from node $3$ to node $1$ are
compositions of these arrows, with relations $x_i y_j - x_j y_i = 0$.

Another example which will be thoroughly dealt with below is $S =
\dP1$, which is $\P^2$ with one point blown up.  Letting $C_1$ be the
exceptional divisor, a complete strongly exceptional collection is $\{
\O, \O(C_1), \O(H), \O(2H)\}$ where $H$ is the hyperplane divisor.  A
slightly more involved analysis shows the quiver to be

\begin{equation}
\begin{xy} <1.0mm,0mm>:
  (0,0)*{\circ}="a",(20,0)*{\circ}="b",(40,0)*{\circ}="c",(60,0)*{\circ}="d",
  (0,-4)*{v_0},(20,-4)*{v_1},(40,-4)*{v_2},(60,-4)*{v_3}
  \ar@{<-}|{a} "a";"b"
  \ar@{<-}@/^3mm/|{b_0} "b";"c"
  \ar@{<-}|{b_1} "b";"c"
  \ar@{<-}@/_8mm/|{c} "a";"c"
  \ar@{<-}@/^3mm/|{d_0} "c";"d"
  \ar@{<-}|{d_1} "c";"d"
  \ar@{<-}@/_3mm/|{d_2} "c";"d"
\end{xy}  
\end{equation}
with the relations $b_0d_1-b_1d_0=0$, $ab_0d_2-cd_0=0$, and $ab_1d_2-cd_1=0$.

Now that we have Bondal's theorem, we can use either $\DC(S)$ or
$\DC(\cmod A)$ to describe branes on $S$.  We will call the branes
that correspond to the representations $L_i$ fractional branes.  Of
course, we are actually interested in branes in $X$, i.e., in the image
of $\DC(S)$ in $\DC(X)$ as noted above.  Using a local model of the
Calabi-Yau X, namely representing it as the total space of the normal
bundle of $S$ in $X$ (which is isomorphic to the canonical bundle) one
can determine the $\Ext$ groups of sheaves in $i_* \DC(S)$ in terms of
the $\Ext$ groups in $\DC(S)$.  Namely, we find using a spectral
sequence argument \cite{ST:braid,KS:Ext, CFIKV:, me:point} that

\begin{equation}
  \Ext^p_X(i_*L_i,i_*L_j) = \Ext^p_S(L_i,L_j) \oplus
  \Ext^{3-p}_S(L_j,L_i).
\end{equation}
In fact it is also true that only one of the direct summands on the
right hand side of the above equation is nonzero, and so we see that
embedding $S$ in $X$ creates new open string degrees of freedom ---
new $\Ext^1$'s corresponding to reversing $\Ext^2$'s in the del Pezzo
quiver.  We can add in arrows corresponding to these new $\Ext^1$'s to
obtain the completed quiver.  For example, the completion of the
Beilinson quiver is

\begin{equation}
\begin{xy} <1.0mm,0mm>:
  (0,0)*{\circ}="a",(20,0)*{\circ}="b",(40,0)*{\circ}="c",
  (0,-4)*{v_0},(20,-4)*{v_1},(40,-4)*{v_2}
  \ar@{-}@/_1mm/|*\dir{<}"a";"b"
  \ar@{-}|*\dir{<}"a";"b"
  \ar@{-}@/^1mm/|*\dir{<}"a";"b"
  \ar@{-}@/_1mm/|*\dir{<}"b";"c"
  \ar@{-}|*\dir{<}"b";"c"
  \ar@{-}@/^1mm/|*\dir{<}"b";"c"
  \ar@{.}@/^4mm/|*\dir{>}"a";"c"
  \ar@{.}@/^5mm/|*\dir{>}"a";"c"
  \ar@{.}@/^6mm/|*\dir{>}"a";"c"
\end{xy}  \label{eq:eg1X}
\end{equation} while the completion of the $\dP1$ quiver becomes

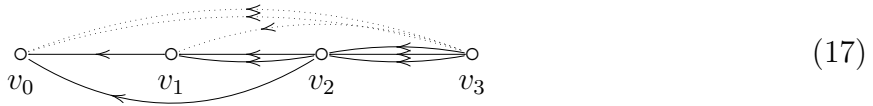
\begin{equation}
\begin{xy} <1.0mm,0mm>:
  (0,0)*{\circ}="a",(20,0)*{\circ}="b",(40,0)*{\circ}="c",(60,0)*{\circ}="d",
  (0,-4)*{v_0},(20,-4)*{v_1},(40,-4)*{v_2},(60,-4)*{v_3}
  \ar@{-}|*\dir{<}"a";"b"
  \ar@{-}@/_1mm/|*\dir{<}"b";"c"
  \ar@{-}|*\dir{<}"b";"c"
  \ar@{-}@/_6.5mm/|(0.3)*\dir{<}"a";"c"
  \ar@{-}@/_1mm/|*\dir{<}"c";"d"
  \ar@{-}|*\dir{<}"c";"d"
  \ar@{-}@/^1mm/|*\dir{<}"c";"d"
  \ar@{.}@/^6mm/|*\dir{<}"a";"d"
  \ar@{.}@/^5mm/|*\dir{<}"a";"d"
  \ar@{.}@/^4mm/|(0.3)*\dir{<}"b";"d"
\end{xy}  \label{eq:dP1X}
\end{equation}

%%%%

%%%%%%%%%%%%%%%%%%%%%%%%%%%%%%%%%%%%%%%%%%%%%%%%%%%%%%%%%

\section{Superpotentials from Topological Field Theory} \label{s:superp}

\subsection{Topological Field Theory}

Having developed and reviewed the requisite mathematical machinery,
let us get to the problem at hand, namely computing superpotentials
for effective dimensionally reduced theories \cite{AK:ainf}.  Our
setting is, as we said, $M \times X$ with $M$ being four dimensional
Minkowski space and $X$ a Calabi-Yau threefold.  In general, the
object is to figure out how to obtain the superpotential for a
specified distribution of space-filling branes --- the case of interest
involves putting $D3$ branes (which look like points in $X$) on a
collapsing del Pezzo cycle $S$ in $X$, but let us for the purpose of
developing some formalism first tackle the case of a single
space-filling and Calabi-Yau filling $D9$ brane, described by a
complex line bundle $E \rightarrow X$ with a hermitian connection.
(By itself this case is unphysical, in a sense, because of anomalies
but the topological field theory makes perfect sense.)

In this case, the massless four dimensional field content is
determined by the Dolbeault cohomology of $X$ valued in $\End(E)$,
$H^{0,q}_{\bar\partial}(X,\End(E))$.  Specifically, the number of
vector bosons is given by $H^{0,0}_{\bar\partial}(X,\End(E)) =
\End(E)$, where by abuse of notation the second term refers to the
space of global sections of $\End(E)$.  We will work with simple line
bundles, for which $\End(E) = \C$.  We could of course also take $N$
copies of the brane, $E^{\oplus N}$, whereby we obtain a $U(N)$ gauge
boson.  Likewise, the number of chiral superfields is given by
$H^{0,1}_{\bar\partial}(X,\End(E))$.  Again, these are in
the adjoint of $U(N)$ when we take the bundle to be $E^{\oplus N}$.

In order to get a term in the tree-level superpotential, we have to
compute a disk diagram with boundary insertions of vertex operators
that correspond to the chiral superfields that appear in that term.
What makes this problem computationally tractable is the fact that
this disk diagram can be computed in a topological theory
\cite{BDLR:Dq}; it is in some sense protected from $\alpha'$
corrections.  Specifically, the open string topological $B$-model on
$X$ with a $D$-brane $E$ has open string spectrum given by $A =
H^{0,q}_{\bar\partial}(X,\End(E))$.  Thus, if we define the disk
correlation functions as:

\begin{equation}
  B_{i_0,i_1,\ldots,i_{k}} = (-1)^{\zeta_1+\zeta_2+\ldots+\zeta_{k-1}}
    \langle \psi_{i_0}\,\psi_{i_1}\,P\int\psi_{i_2}^{(1)}\,
    \int\psi_{i_3}^{(1)}\ldots\int\psi_{i_{k-1}}^{(1)}\,
    \psi_{i_k}\rangle,  \label{eq:Bdef}
\end{equation}

Here the $\psi_{i_m}$ are vertex operators of ghost number one,
i.e., they correspond to states in $H^{0,1}_{\bar\partial}(X,\End(E))$.
If we let $Z_i$ be the effective four dimensional superfield
corresponding to the open string mode $\psi_i$, then the
superpotential is

\begin{equation}
  W = \Tr\left(\sum_{k=2}^\infty\,\sum_{i_0,i_1,\ldots,i_{k}}\!
    \frac{B_{i_0,i_1,\ldots,i_{k}}}{k+1}Z_{i_0}Z_{i_1}\ldots Z_{i_{k}}\right).
    \label{eq:Wdef}
\end{equation}

What have we accomplished by reducing the problem to a computation in
a topological sigma model?  Heuristically, the situation is as follows
\cite{AK:ainf}: we have, by reducing to the topological theory,
essentially gotten rid of the higher mode excitations of the string.
Hence the disk diagram we want is really a sum of Feynman diagrams in
a field theory, called holomorphic Chern-Simons theory.  Because big
Feynman diagrams can be built from smaller ones, we obtain from this
way of looking at things combinatorial relations among the
correlators, called $A_\infty$ relations, and it turns out that these
determine the correlators uniquely (up to field redefinition).  In
fact, the $A_\infty$ relations give a specific algorithm for
generating the correlators, and this algorithm generalizes to a more
general setting where $D$-branes are represented as elements of the
derived category of coherent sheaves.

We now proceed to flesh out the above heuristic and describe the
algorithm in detail.  First, we briefly review some mathematical
background on $A_\infty$ products.

\subsection{$A_\infty$ structure}

Given a graded vector space $B$, such as the Dolbeault complex graded
by $q$ defined above, an $A_\infty$ structure on $B$ is defined as a
series of products $m_k$, $k \geq 1$, of degree $2-k$

\begin{equation}
  m_k: B^{\otimes k}\to B,
\end{equation} which satisfy the $A_\infty$ constraints:

\begin{equation}
  \sum_{r+s+t=n} (-1)^{r+st}m_u(\id^{\otimes r}\otimes m_s\otimes \id^{\otimes
  t})=0,   \label{eq:Ainf}
\end{equation}
for any $n>0$, where $u=n+1-s$.  Here we assume the usual sign rule 

\begin{equation}
(f\otimes g)(a\otimes b) = (-1)^{|g|.|a|}f(a)\otimes g(b) \label{eq:sign_rule}
\end{equation} when moving arguments past operators.

The $A_\infty$ products can actually be rephrased in terms of a
differential acting on a certain space, with the complicated and
unnatural looking relations between them being just the condition that
the differential squares to zero \cite{AK:ainf}.  We will not pursue
this interpretation here however, except to note that it is useful to
consider maps between spaces that commute with the differential.  In
terms of the $A_\infty$ products, such a map between two spaces $B$
and $B'$ is described as an $A_\infty$ morphism, which is to say it is
given by a series of maps

\begin{equation} f_k: B^{\otimes k}\to B',\end{equation} for $k \geq 1$, which satisfy

\begin{equation}
  \sum_{r+s+t=n}(-1)^{r+st} f_u(\id^{\otimes r}\otimes m_s\otimes \id^{\otimes
  t}) =
  \sum_{\genfrac{}{}{0pt}{}{1\leq r\leq n}{i_1+\ldots+i_r=n}}
     \!\!(-1)^q m_r(f_{i_1}\otimes
  f_{i_2}\otimes\cdots\otimes f_{i_r}), \label{eq:Amorph}
\end{equation}
for any $n>0$, $u=n+1-s$, 
and $q = (r-1)(i_1-1) + (r-2)(i_2-1) + \ldots + (i_{r-1}-1)$.

Now note that the $A_\infty$ relations give $m_1 \cdot m_1 = 0$, so
that $B$ has the structure of a graded differential complex, and we
can take cohomology $H^*(B)$.  We now come to a theorem that forms the
basis for the computational tractability of our results.  Let $B$ be
as above, except assume that all products $m_k$ are zero for $k \geq
3$ --- this structure is called a differential graded algebra (dga).
Given an embedding $i: H^*(B) \to B$ Kadeishvili \cite{Kad:Ainf} shows
that we may define an $A_\infty$ structure on $H^*(B)$ that has $m_1 =
0$ and an $A_\infty$ morphism $f$ from $H^*(B)$ to $B$ with $f_1$
equal to the embedding $i$.  Furthermore if $B$ and $B'$ are
quasi-isomorphic dga's (that is, there is a map from one to the other
that induces an isomorphism on cohomology) then the two Kadeishvili
$A_\infty$ structures on $H^*(B)$ and $H^*(B')$ are
$A_\infty$-isomorphic.

There is in fact a well defined algorithm for determining the
$A_\infty$ products of Kadeishvili's theorem.  The above condition for
an $A_\infty$ morphism, for the case $n=2$, gives

\begin{equation}
im_2 = (i\cdot i) + df_2.
\end{equation}
The cohomology class of the right hand side of the above equation is
just that of $i \cdot i$ and hence $m_2$ is uniquely determined.
Therefore $df_2$ is also uniquely determined, and we can invert $d$ to
obtain a (non-unique) choice of $f_2$.  Now putting $n=3$ we have

\begin{equation}
  im_3 = f_2(\id\otimes m_2)-f_2(m_2\otimes\id) + (i\cdot f_2)
  -(f_2\cdot i) + d f_3. \label{eq:m3}
\end{equation}
Once again, this equation uniquely determines $m_3$ and $df_3$, and
allows us to make a choice of $f_3$.  Continuing in this way, it is
apparent that all $A_\infty$ products can be determined.  The
ambiguity in the choice of $f_k$ reflects the ambiguity in the
uniqueness clause of the above theorem.

\subsection {Holomorphic Chern-Simons Theory}

\def\fa{\mathsf{A}}

The field theory that the topological $B$-model on $X$ reduces to is
holomorphic Chern-Simons theory:

\begin{equation}
S = \int_X \Tr\left(\fa\wedge \bar\partial\fa + \ff23\fa\wedge\fa\wedge
\fa\right)\wedge\Omega,  \label{eq:hCS}
\end{equation}
where the $\fa$ is a $(0,1)$-form on $X$ taking values in $\End(E)$,
and $\Omega$ is a holomorphic $(3,0)$-form on $X$.  As mentioned
above, computation of the disk correlator in holomorphic Chern-Simons
theory reduces to a sum of Feynman diagrams (this reduction can be
seen explicitly as localization of the supersymmetric path integral on
Feynman fat-graph configurations arising from instantons at infinity,
see \cite{W:CS}).  The combinatorial relations which the Feynman
diagram picture gives rise to are precisely the $A_\infty$ relations.
To make a rigorous statement, first define a trace map

\begin{equation}
  \gamma(a) = \int_X \Tr(a)\wedge\Omega,
\end{equation}
$\gamma$ is a degree $-3$ map in the sense that only $(0,3)$-forms $a$
have nonzero trace.  Define $m_1$ to be $\bar\partial$ and $m_2$ to be
the wedge product together with composition in $\End(E)$ --- these
give the Dolbeault complex the structure of a dga.  The embedding of
$\bar\partial$ cohomology into the Dolbeault complex by harmonic forms
then gives via Kadeishvili an $A_\infty$ structure to
$H^{0,q}_{\bar\partial}(X,\End(E))$.  The correlation functions can
then be written \cite{HLW:Ainf}:

\begin{equation}
  B_{i_0,i_1,\ldots,i_{k}} =
    \gamma\left(m_2\left(m_k(\psi_{i_0},\psi_{i_1},\ldots,
     \psi_{i_{k-1}}),\psi_{i_k}\right)\right),
\end{equation}They satisfy the cyclicity property \cite{HLW:Ainf}: \begin{equation}
   B_{i_0,i_1,\ldots,i_{k}} =
    (-1)^{\zeta_k(\zeta_0+\zeta_1+\ldots+\zeta_{k-1})}
    B_{i_k,i_0,i_1,\ldots,i_{k-1}}. \label{eq:cyc}
\end{equation}which will be important to us later.

Up to now we have been dealing with a single Calabi-Yau filling
$D$-brane.  The advantage of working in the above framework is that it
extends easily to more general $D$-brane configurations.  For example,
(still for a single brane $E$) we may replace the Dolbeault complex by
a \v Cech complex, thereby turning a difficult problem in analysis,
namely inverting $\bar\partial$, into a more manageable combinatorial
one.  The uniqueness theorem above guarantees that the two $A_\infty$
structures obtained are $A_\infty$-isomorphic.  We could also use an
injective resolution of a sheaf instead of the \v Cech complex, and by
appropriate abstraction reframe the entire discussion in terms of
$\DC(X)$.  In fact, for now the most convenient complex for us to use
is a hybrid of the \v Cech complex and that obtained from locally free
resolutions (i.e., resolutions by vector bundles).  Specifically, we
claim that, for a $D$-brane represented in the derived category by the
locally free resolution

\begin{equation}
\cE^\bullet = \left(\xymatrix@1@C=15mm{
\ldots\ar[r]^-{\ddd_{n-2}}&\cE^{n-1}\ar[r]^-{\ddd_{n-1}}&
    \cE^{n}\ar[r]^-{\ddd_n}&\cE^{n+1}
    \ar[r]^-{\ddd_{n+1}}&\ldots}\right).
\end{equation}
the following complex has cohomology that gives the correct open
string spectrum for the brane and induced $A_\infty$ structure that
gives rise to the correct superpotential:
\begin{equation}
\xymatrix@1{
\ldots\ar[r]&\cB^{n-1}\ar[r]&\cB^{n}\ar[r]&\cB^{n+1}\ar[r]&\ldots,
}  \label{eq:B1}
\end{equation}
where
\begin{equation}
\begin{split}
  \cB^n &= \bigoplus_{p+q=n} \cB^{p,q}\\
  \cB^{p,q} &= \check{C}^p
    \left(\mf{U},\sHom^q(\cE^\bullet,\cE^\bullet)\right). 
\end{split}
\label{eq:cB}
\end{equation}
This is shown in \cite{AK:ainf}.  Two points need to be made here.
First, the differential in (\ref{eq:B1}) is $d=\delta+(-1)^{p}\mf{d}_{q}$,
with $\delta$ the \v Cech differential and $\mf{d}_{q}$ given by

\begin{equation}
  \mf{d}_n f_{n,p} = \ddd_{p+n}\circ f_{n,p} - 
    (-1)^n f_{p+1,n}\circ \ddd_{p}.  \label{eq:mfd}
\end{equation} 
where $\sum_p f_{n,p}$, with $f_{n,p}:\cE^p\to\cE^{p+n}$, is an element of
\begin{equation}
   \sHom^n(\cE^\bullet,\cE^\bullet) = \bigoplus_p \sHom(\cE^p,\cE^{p+n}).
\end{equation}
Second, to rigorously show that $\cB$ indeed reproduces the correct
spectrum and superpotential is non-trivial and requires an analysis of
elements of the derived category as boundary states of the worldsheet
theory \cite{me:TASI-D}.

\section{Superpotentials for del Pezzo singularities} \label{sdP:sdP}
\subsection{Moduli Spaces}   \label{ss:mod}

Before launching into a more rigorous discussion, let us first
consider a heuristic argument that will lead to a conjecture
for the form of the superpotential.

First we quickly review the connection of the mathematics of quivers
to the physics of D-branes and stability. One should view a quiver as
representing a decay of a D-brane. The nodes in the quiver correspond
to the decay products, i.e., the so-called ``fractional branes'' and
the arrows correspond to open strings between these decay
products. The D-brane we are particularly concerned with is the
3-brane corresponding to a point in $X$.

At the instant of decay, the open strings corresponding to the arrows
should be exactly massless. In the case of B-branes, these masses are
a function of the complexified K\"ahler form $B+iJ$. Here we assume
that this masslessness occurs precisely when the del Pezzo surface is
collapsed to a point. This assumption was justified in
\cite{me:theta}.

If one moves away from the critical point where the open strings are
massless, then the D-brane may become stable or unstable with respect
to the decay. If we deform the K\"ahler form to some generic value to
give a nonzero size the del Pezzo surface (and all the curves within
it) then we expect the 3-brane to be stable.

In this resolution one may compute the moduli space of the 3-brane,
which should, of course, yield $X$ itself. We need not concern
ourselves with the details of this process but we note the
following. For more details we refer to
\cite{DFR:orbifold,DG:fracM,me:TASI-D,me:theta}. The moduli space of
3-branes is essentially given by the moduli space of representations
of the quiver.  One takes all possible quiver representations which
satisfy ``$\theta$-stability'' and then divides by a gauge equivalence.

Physically this moduli space is given by the moduli space of chiral
fields (given by the matrices associated to arrows in the quiver)
corresponding to classical solutions of the field theory divided by
gauge equivalence. Importantly for us, this must mean that {\em the
superpotential imposes conditions on the chiral fields equivalent to
the relations in the quiver.}

In other words, finding the critical points of the superpotential
must be equivalent to imposing the quiver relations. This leads to an
obvious proposition for the superpotential. Let $A_i$ be the chiral
fields in the worldvolume gauge theory associated to the arrows in the
(non-completed) quiver associated to a del Pezzo surface. The
relations will be denoted $r_k(A_1,A_2,\ldots)=0$, where $r_k$ is some
polynomial. We know from section \ref{ss:shvs} that each $r_k$ is
associated to some arrow in the completed quiver, and so some chiral
field $R_k$. It is believed (see \cite{DFR:orbifold,me:TASI-D}, for
example) that in terms of the moduli space, setting all $R_k$ equal to
zero amounts to restricting the 3-brane is be on the del Pezzo surface $S$
itself. Giving nonzero expectation values to the $R_k$ fields moves
the 3-brane off $S$.

If the superpotential is given by
\begin{equation}
   W = \sum_k R_kr_k(A_1,A_2,\ldots),
\end{equation}
then, on $S$, the equations of motion will yield precisely the correct
constraints, at least for 3-branes on $S$. This, therefore, is our
conjectured form for the superpotential.

\subsection {Quiver Gauge Theories for del Pezzos}

Let us now consider more systematically what happens
when we put a $D3$-brane on a shrinking del Pezzo cycle $S$ in a
Calabi-Yau $X$.  Now, every BPS space-filling brane corresponds to a
topological brane on $X$, but not vice versa.  A point on $X$ is
always a valid topological brane; when $S$ is of finite size, the
$D3$-brane will be located on a smooth point of $S$ and as we said we expect it
to be BPS.  On the other hand, when $S$ shrinks, one can argue (see
e.g. \cite{AM:delP}) that the $D3$ is marginally stable against decay
into the fractional branes introduced earlier.  We think of these
fractional branes as wrapping $S$ --- when $S$ shrinks the point-like
$D3$ is allowed to marginally decay to them.

To get a precise description of these fractional branes, we recall
that they correspond to the representations $L_i$, which have
resolutions in terms of the projective representations $P_i$.  If we
replace the $P_i$ by the corresponding elements of the strongly
exceptional collection (which are all vector bundles in the cases we
consider) and use the equivalence between derived categories, we
obtain locally free resolutions of the fractional branes.  For example
in the case of $\P^2$ and the Beilinson quiver, $L_2$ is represented
in $D(S)$ as

\begin{equation}
\xymatrix@1{\ldots\ar[r]&0\ar[r]&\cF_0^{\oplus 3}\ar[r]^-{\ddd_{-3}}&\cF_1^{\oplus 3}\ar[r]^-{\ddd_{-2}}&\cF_2\ar[r]^-{\ddd{-1}}&0\ar[r]&\ldots}
\end{equation} 
The maps in the above complex are determined by the corresponding maps
in the quiver resolution of $L_2$, using the fact that $\Hom(\cF_i,
\cF_j)$ and $\Hom(P_i, P_j)$ are naturally isomorphic.  It turns out
that a D3-brane decays into a collection of fractional branes with each
fractional brane occurring dim $\cF_i$ times \cite{AM:delP}.  The quiver
gauge theory for a $\dP k$ will thus have $k+3$ gauge groups,
corresponding to each of the $L_i$, and massless matter in the
bifundamental from strings stretching between $L_i$ and $L_j$.  The
proper setting for discussing the homological structure in this
context is an $A_\infty$ category, but we do not need to get so
abstract.  We simply let $\cM^\bullet$ be the direct sum of the
locally free resolutions of the $L_i$, and use it as the starting
point for the $A_\infty$ computations.  We remind the reader that
$\cM^\bullet$ is a locally free resolution of sheaves over $S$, not
$X$.

It will be convenient to represent $X$ as the total space of the
normal bundle $N$ of $S$.  Because $S$ is a del Pezzo, $N \cong
K_S$\cite{AM:delP}.  We are allowed to take this limit in K\"ahler
moduli space because the tree level superpotential does not depend on
K\"ahler parameters.

\subsection{From Branes on X to Branes on S}

The actual superpotential is computed from the $A_\infty$ products of
the \v Cech complex associated to branes not on $S$, but on $X$,
i.e., not from $\cB^{\bullet, \bullet}$, but rather from the associated
complex obtained from considering all the sheaves as embedded in $X$.
The goal of this section is to show that the computation of $A_\infty$
products associated to branes on $X$ essentially reduces to the
computation on $S$.  Specifically, we recall that (as we will see in
greater detail below) $\Ext^1$'s of sheaves on $S$ considered as
sheaves on $X$ include all the $\Ext^1$'s of the sheaves on $S$ plus
some extra $\Ext^1$'s, which, after a reversal of arrows, correspond
to $\Ext^2$'s of the sheaves on $S$.  We will prove that there is a
choice of $A_\infty$ structure over $X$ such that all $A_\infty$
products that contain more than one of the ``extra'' $\Ext^1$'s
vanish.  The products that contain no ``extra'' $\Ext^1$'s are the
same as they were over $S$, and the ones with one ``extra'' $\Ext^1$
are determined uniquely by the requirement of cyclicity.  This,
together with the fact that there are no cycles in del Pezzo quivers,
will show that the superpotential is linear in the ``extra''
$\Ext^1$s, with these $\Ext^1$s multiplying terms that are just the
quiver relations.

To proceed with the proof, let $\pi: E \rightarrow S$ be the
projection from the total space of $K_S$ to the del Pezzo $S$.  We
have a canonical section $\O \rightarrow \pi^*K_S$, given as follows:
to each point in $E$ we tautologically associate a point of $S$ and an
element of the fiber of $K_S$ over that point; this element can be
viewed as an element of the fiber of $\pi^*K_S$ over the original
point in $E$.  Dualizing, we get a canonical map which fits into an
exact sequence of sheaves 
\begin{equation} 0 \rightarrow
\pi^*({K_S}^*) \rightarrow \O \rightarrow \O_S \rightarrow 0.
\end{equation} 
and we can think of the first two terms as a locally free resolution
of $\O_S$.  The key point now is to tensor the resolution
$\cM^\bullet$ with the above resolution of $\O_S$ in order to obtain a
locally free resolution of $i_* M$:

\begin{equation}
\begin{xy}
\xymatrix{
\ar[r]&{\pi^{-1}}\cM^0\ar[r]&{\pi^{-1}}\cM^1\ar[r]&{\pi^{-1}}\cM^2\ar[r]&\\
\ar[r]&\pi^*({K_S}^*) \otimes {\pi^{-1}}
\cM^0\ar[r]^-{\ddd^0}\ar[u]_-{\ddd^1}&\pi^*({K_S}^*) 
\otimes {\pi^{-1}}\cM^1\ar[r]\ar[u]&\pi^*({K_S}^*) \otimes {\pi^{-1}} 
\cM^2\ar[r]\ar[u]&
}
\save="x"!LD+<-3mm,0pt>;"x"!RD+<0pt,0pt>**\dir{-}?>*\dir{>}\restore
\save="x"!LD+<0pt,-3mm>;"x"!LU+<0pt,-2mm>**\dir{-}?>*\dir{>}\restore
\save!CD+<0mm,-4mm>*{i}\restore
\save!CL+<-3mm,0mm>*{j}\restore
\end{xy}
\label{eq:locfreeM}
\end{equation} 
Collapsing the above double complex along the diagonal we get a free
resolution, and the associated spectral sequence, which collapses at
the $E_2$ term, shows that it is in fact a resolution of $i_* \cM$.
This can also be viewed as the Cone construction \cite{me:TASI-D}.  We
will choose to retain the bi-grading, so let us represent the above
resolution as $\cM ^{\bullet, \bullet}$ (the first index corresponds
to the index of $\cM^\bullet$, and the second is either $0$, for
$\pi^{-1} \cM^{\bullet}$, or $1$, for $\pi^*({K_S}^*) \otimes
{\pi^{-1}} \cM^{\bullet}$).  We now define

\begin{equation}
\cC^{p,i,j} = \check{C}^p\left(\pi^{-1} 
\mf{U},\sHom^{i,j}(\cM^{\bullet, \bullet},\cM^{\bullet, \bullet})\right)
\end{equation}

Here $\mf{U}$ is an affine open cover of $S$, and hence $\pi^{-1}
\mf{U}$ is an affine open cover of $E$.  The complex $\cC$ is central
in our analysis.  There are several differentials we define on $\cC$.
First of all, in the locally free resolution $\cM^{\bullet, \bullet}$
label the differentials that increase the first index by
${\ddd^{0}}_{i,j}$ and that which increases the second index by
${\ddd^{1}}_{i,j}$.  The combination $\ddd_{i,j} = \ddd^{0}_{i,j} +
{(-1)}^i \ddd^{1}_{i,j}$ is the standard differential associated to
the locally free resolution of $\cM$ over $E$.  Now, given a section
of $\sHom^{i,j}(\cM^\bullet,\cM^\bullet)$ over an open set $U$, i.e., a
section of $\bigoplus_{p,q} \Hom_U (\cM^{p,q}, \cM^{p+i, q+j})$, we
can denote it by 
\begin{equation} \sum_{p,q} f^{i,j}_{p,q},
\end{equation}
where 
\begin{equation}
f^{i,j}_{p,q}: \cM^{p,q}(U)
\rightarrow \cM^{p+i,q+j}(U)
\end{equation}
Then we can define differentials

\begin{equation}
{\mf{d}^{0}}_{i,j} f^{i,j}_{p,q} = {\ddd^{0}}_{i,j} f^{i,j}_{p,q} - 
    (-1)^{(i+j)} f^{i,j}_{p+1,q} {\ddd^{0}}_{i,j}
\end{equation}

\begin{equation}
{\mf{d}^{1}}_{i,j} f^{i,j}_{p,q} = {\ddd^{1}}_{i,j} f^{i,j}_{p,q} - 
    (-1)^{(i+j)} f^{i,j}_{p,q+1} {\ddd^{1}}_{i,j}.
\end{equation}

Finally, we have the \v Cech differential $\delta$, and we define the total differential on $\cC$ by $d = \delta + {(-1)}^p \left( \mf{d}^{0} + (-1)^i \mf{d}^{1} \right)$.  Now, the sum $\mf{d}^{0} + (-1)^i \mf{d}^{1}$ is the standard differential associated to the locally free resolution of $\cM$ over $E$, so that collapsing on the $(i,j)$ indices yields the double complex in \cite{AK:ainf}, showing that $\cC$ does indeed correctly compute the $A_\infty$ products.

To actually get a handle on determining the $A_\infty$ algebra, it is useful to collapse $\cC$ in a different way and leverage our knowledge of the $A_\infty$ structure for sheaves on the del Pezzo $S$.  Specifically, let us collapse the complex on the $(p,i)$ indices: \begin{equation}
\cD^{q,j} = \bigoplus_{p+i=q} \cC^{p,i,j}
\end{equation} $\cD^{\bullet,\bullet}$ is a double complex with anticommuting differentials $d_0 = \delta + {(-1)}^p \mf{d}^{0}$, which increases the first index, and $d_1 = {(-1)}^q \mf{d}^{1}$, which increases the second one, that add up to $d$.  The desired cohomology is computed using a spectral sequence associated to this double complex, which by arguments of \cite{AM:delP} degenerates at the $E_2$ term to give: \begin{equation}
\begin{xy}
\xymatrix@C=4mm@R=3mm{
  0&0&0&0\\
  \Ext^{-1}_S(M,M \otimes K_S)&\Ext^0_S(M,M \otimes K_S)&\Ext^1_S(M,M \otimes K_S)&\Ext^0_S(M,M \otimes K_S)\\
  \Ext^{-1}_S(M,M)&\Ext^0_S(M,M)&\Ext^1_S(M,M)&\Ext^2_S(M,M)\\
} 
% +<0pt,0pt> collapses box.
\save="x"!LD+<-3mm,0pt>;"x"!RD+<0pt,0pt>**\dir{-}?>*\dir{>}\restore
\save="x"!LD+<39mm,-3mm>;"x"!LU+<39mm,-2mm>**\dir{-}?>*\dir{>}\restore
\save!CD+<0mm,-4mm>*{q}\restore
\save!UL+<36mm,-5mm>*{j}\restore
\end{xy}  \label{eq:StoX}
\end{equation} (In our exposition the $E_1$ term is given by taking cohomology with respect to $d_1$).  Serre duality shows that $\Ext^i_S(M,M) \cong \Ext^{2-i}_S(M,M \otimes K_S)$, so that

\begin{equation}
\Ext^1_X (i_* M, i_* M) \cong \Ext^1_S(M,M) \oplus \Ext^2_S(M,M).
\end{equation}  The two terms on the right correspond to the $\Ext^1$s and ``extra'' $\Ext^1$s, respectively.

Now, the bottom row in the above diagram reproduces the cohomology of
the complex $\cB^{p,i}$ --- the complex associated to branes on $S$
rather than $X$. In fact, we may naturally embed $\cB^{p,i}$ in

\begin{equation}
\check{C}^p\left(\pi^{-1} \mf{U}, \sHom^i(\pi^{-1} \cM^\bullet, \pi^{-1} \cM^\bullet)\right).
\end{equation}This complex, in turn, can be viewed as a sub-complex of $\cC^{p,i,0}$.  To see why, note that, from (\ref{eq:locfreeM}), a section of $\sHom^i(\pi^{-1} \cM^\bullet, \pi^{-1} \cM^\bullet)$ determines a section of $\sHom^{i,0}(\cM^{\bullet, \bullet},\cM^{\bullet, \bullet})$; basically it gives directly the maps among the sheaves in the upper row of (\ref{eq:locfreeM}), and, taking the identity map on $\pi^*({K_S}^*) \otimes {\pi^{-1}} \cM^{\bullet}$, it determines the maps for the sheaves in the lower row as well.  Also, from (\ref{eq:StoX}), we see that the composition of these embeddings induces an isomorphism from the cohomology of $\cB^{p,i}$ to the $j=0$ part of the cohomology of $\cC^{p,i,j}$.

Now, in order to carry out the $A_\infty$ procedure we must choose representatives for all cohomology classes in $\cC^{p,i,j}$.  The upshot of the construction in the previous paragraph is that it gives us a natural choice of representatives of the $j=0$ part of the cohomology; in fact, it shows that the $A_\infty$ products of these $j=0$ cohomology classes are exactly the same as those in $\cB^{p,i}$.  In other words, for the $j=0$ generators the $A_\infty$ products are just those defined over $S$.  This accomplishes part of our goal of reducing the computation over $X$ to a computation over $S$; to finish we have to deal with products that may contain some $j=1$ generators.

The $j=1$ cohomology generators are the ones that contribute the ``extra'' $\Ext^1$s.  To carry out the $A_\infty$ procedure, we must pick representatives of their cohomology classes.  We choose these to be homogeneous of $j$ degree $1$, or, in other words, to lie in

\begin{equation}
\check{C}^p\left(\pi^{-1} \mf{U}, \sHom^i(\pi^{-1} \cM^\bullet \otimes \pi^{-1} K_S, \pi^{-1} \cM^\bullet)\right)
\end{equation}Clearly this is the most natural choice, though it should be pointed out that we could have done something stupid and chosen the generator to have a nonzero (exact) $j=0$ part, for example.  The advantage of having homogeneous generators is that their products are homogeneous as well, and so vanish if they have $j > 1$.

Now we claim that any $m_k$ that contains more than one $j=1$ generator vanishes.  The naive argument would invoke the $j$ grading and the fact that there are no elements in $\cC^{p,i,j}$ with $j>1$.  The obvious flaw is the fact that $m_k$ does not respect the overall grading --- it in fact has degree $2-k$.  Thus we have to be more careful.  We claim that although $m_k$ has nonzero degree with respect to the overall grading $p+i+j$, through a careful choice of $f_k$, which we now construct, we can make $m_k$ respect the $j$ grading.  The claim at the top of this paragraph then immediately follows.

We show by induction that all the $f_k$ and $m_k$ respect the $j$
grading.  Clearly $f_1$ and $m_1$ respect the $j$ grading. Suppose
that this is also true for all $k \leq n$.  We have for all $n+1$
(\ref{eq:Amorph}), which can be rewritten as an equation determining
$m_{n+1}$ in terms of the lower $m_k$ and $f_k$ (for $n+1=3$, for
example, this is (\ref{eq:m3})).  So we immediately see that it's true
for $m_{n+1}$.  We now deal with $f_{n+1}$.  We suppress its
arguments, but everywhere below $f_{n+1}$ and $df_{n+1}$ will stand
for $f_{n+1}$ and $df_{n+1}$ applied to their arguments.  Now equation
(\ref{eq:Amorph}) again gives $df_{n+1}$ as an expression in terms of
$m_{n+1}$ and the $m_k$ and $f_k$ for $k \leq n$.  We have to make a
choice of $f_{n+1}$ that respects the $j$ grading, i.e., we want
$f_{n+1}$ to have the same $j$ degree as $df_{n+1}$.  Now, the case
when all the arguments have $j=0$ has been discussed above and clearly
we have already defined $f_{n+1}$ to have $j=0$.  When more than one
argument has $j=1$ then, because all the terms in the expression for
$df_{n+1}$ are homogeneous, we have $df_{n+1} = 0$, so that we can
choose $f_{n+1} = 0$.  The nontrivial case is when exactly one
argument has $j=1$.  In that case, $df_{n+1}$ is homogeneous of $j$
degree $1$ and hence lies in the sub-complex

\begin{equation}
\cC^{p,i,1} = \check{C}^p\left(\pi^{-1} \mf{U}, \sHom^i(\pi^{-1} \cM^\bullet \otimes \pi^{-1} K_S, \pi^{-1} \cM^\bullet)\right).
\end{equation}The crucial point is now that the embedding $\cC^{p,i,1} \subset \cC^{p,i,j}$ induces an injection in cohomology.  This can easily be seen from the spectral sequence (\ref{eq:StoX}) --- the cohomology of $\cC^{p,i,1}$ reproduces precisely the upper, $j=1$ row in the diagram.  Therefore $df_{n+1}$ is exact not only in $\cC^{p,i,j}$ but also in the sub-complex $\cC^{p,i,1}$.  Therefore we can choose $f_{n+1}$ to be in $\cC^{p,i,1}$, so that it will have $j=1$.  Thus we see that we can always choose $f_{n+1}$ to respect the $j$ grading.  This completes the inductive step.

Together with the cyclicity property this determines all the
$A_\infty$ products in $\cC^{p,i,j}$ in terms of those over $S$.  To
restate, we have the original $A_\infty$ algebra reproduced when all
the arguments are the original $\Ext^1$'s (i.e., have $j=0$), any
product that involves more than one ``extra'' $\Ext^1$ (i.e., one that
has $j=1$) must vanish, while any product that contains exactly one
``extra'' $\Ext^1$ is determined uniquely by requiring it to reproduce
correlators that obey the cyclicity property (\ref{eq:cyc}).  Having
accomplished the reduction and thus shown that the superpotential is
linear in the ``extra'' $\Ext^1$s, we now determine the $A_\infty$
products over $S$ and relate them to the quiver relations.

\section{$A_\infty$ relations and quivers} \label{quiv:quiv}

We must determine the $A_\infty$ products over $S$, i.e., those of
$\cB^{\bullet, \bullet}$, defined in (\ref{eq:cB}).  We know by
Bondal's theorem that $\DC(S)$ is equivalent to $\DC(\cmod A)$, where
$A$ is the path algebra of the associated quiver.  The operational
version of this equivalence that will suffice for us is as follows.
First, construct a complex of quiver representations $M^\bullet$ by
summing the projective resolutions of the $L_i$.  In the usual way it
gives rise to the graded dga $\End(M)^\bullet$.  There is a natural
map of this complex into $\cB$ given by interpreting maps in $\End(M)$
as global sections of the $\Hom(\cF_i, \cF_j)$ and mapping them to \v
Cech $0$-cochains.  Because these $0$-cochains are global sections,
they are annihilated by the \v Cech part of the differential, and one
thus quickly sees that this map is a map of dga's.  In fact, (the
derivation of) Bondal's theorem shows that it is a quasi-isomorphism.
This allows us to apply Kadeishvili's theorem and compute the
$A_\infty$ structure of $\cB$ in the quiver dga $\End(M)^\bullet$.

We will see that we obtain a form of the superpotential exactly as
conjectured in section \ref{ss:mod}.

\subsection{A simple example}

We start with the Beilinson quiver, corresponding to $S = \P^2$:

\begin{equation}
\begin{xy} <1.0mm,0mm>:
  (0,0)*{\circ}="a",(20,0)*{\circ}="b",(40,0)*{\circ}="c",
  (0,-4)*{v_0},(20,-4)*{v_1},(40,-4)*{v_2}
  \ar@{<-}@/^3mm/|{a_0} "a";"b"
  \ar@{<-}|{a_1} "a";"b"
  \ar@{<-}@/_3mm/|{a_2} "a";"b"
  \ar@{<-}@/^3mm/|{b_0} "b";"c"
  \ar@{<-}|{b_1} "b";"c"
  \ar@{<-}@/_3mm/|{b_2} "b";"c"
\end{xy}  
\end{equation} with relations $a_\alpha b_\beta = a_\beta b_\alpha$.  We recall that we have the projective resolutions: \begin{equation}
\xymatrix@R=3mm{  
&&0\ar[r]&P_0\ar[r]&L_0\ar[r]&0\\
&0\ar[r]&P_0^{\oplus3}\ar[r]&P_1\ar[r]&L_1\ar[r]&0\\
0\ar[r]&P_0^{\oplus3}\ar[r]&P_1^{\oplus3}\ar[r]&P_2\ar[r]&L_2\ar[r]&0\\
}
\end{equation} and that \begin{equation}
\begin{split}
  \dim\Ext^1(L_i,L_j) &= n_{ij}\\
  \dim\Ext^2(L_i,L_j) &= r_{ij},
\end{split}
\end{equation}
where $n_{ij}$ counts arrows and $r_{ij}$ counts relations.

We start by choosing specific generators of the $\Ext^i$.  Recalling that the $\Ext^\bullet$ can be represented as morphisms between resolutions of the $L_i$, we can choose the three generators $\fd a_i$ of $\Ext^1(L_1,L_0)$ to be \begin{equation}
\xymatrix{
P_0^{\oplus3}\ar[r]\ar[d]^-{\pi_i}&P_1\\P_0}
\end{equation} where $\pi_i$ is projection on the $i$'th factor.  As far as the generators $\fd b_i$ of $\Ext^1(L_2,L_1)$ we have $\fd b_0$ represented as \begin{equation}
\xymatrix{
P_0^{\oplus3}\ar[r]\ar[d]^-{\left(\begin{smallmatrix}0&0&0\\0&0&-1\\0&1&0\end{smallmatrix}\right)}&
P_1^{\oplus3}\ar[r]\ar[d]^-{\left(\begin{smallmatrix}1&0&0\end{smallmatrix}\right)}&
P_2\\
P_0^{\oplus3}\ar[r]&
P_1&}
\end{equation} and the other $\fd b_i$ represented similarly.  We also have the relations $\fd r_i$ in $\Ext^2(L_3,L_2)$, represented by \begin{equation}
\xymatrix{
P_0^{\oplus3}\ar[r]\ar[d]^-{\pi_i}&P_1^{\oplus3}\ar[r]&P_2\\
P_0&&}
\end{equation} Clearly, the only potentially nontrivial products are $m_2({\fd a_j}, {\fd b_i})$, and one easily sees by composing the representatives for $\fd a_j$ and $\fd b_i$ that $m_2({\fd a_j}, {\fd b_i}) = \epsilon^{ijk} r_k$.  This gives rise to the superpotential

\begin{equation}
W=\epsilon^{ijk} A_i B_j R_k
\end{equation}
which is the correct superpotential for this quiver gauge theory on
the orbifold $\C^3/\Z_3$ \cite{DM:qiv}.  Again, the
$A_i$ and $B_j$ are massless moduli corresponding to the internal
structure of the shrinking cycle $\P^2$, while $R_k$ is the modulus
that corresponds to moving the $D3$-brane off the singularity.  We
note that the superpotential is of the desired form, linear in the
``extra'' $\Ext^1$s $R_i$, which multiply the relations.  To write it
out explicitly, we have

\begin{equation}
W= (A_0 B_1 - B_1 A_0) R_2 + (A_1 B_2 - B_2 A_1) R_0 + (A_2 B_0 - B_0 A_2) R_1
\end{equation}

\subsection{del Pezzo 1}

Let us consider the quiver associated to $\dP 1$.  It is:

\begin{equation}
\begin{xy} <1.0mm,0mm>:
  (0,0)*{\circ}="a",(20,0)*{\circ}="b",(40,0)*{\circ}="c",(60,0)*{\circ}="d",
  (0,-4)*{v_0},(20,-4)*{v_1},(40,-4)*{v_2},(60,-4)*{v_3}
  \ar@{<-}|{a} "a";"b"
  \ar@{<-}@/^3mm/|{b_0} "b";"c"
  \ar@{<-}|{b_1} "b";"c"
  \ar@{<-}@/_8mm/|{c} "a";"c"
  \ar@{<-}@/^3mm/|{d_0} "c";"d"
  \ar@{<-}|{d_1} "c";"d"
  \ar@{<-}@/_3mm/|{d_2} "c";"d"
\end{xy}  
\end{equation}
subject to the relations $r_0 = b_0d_1-b_1d_0$, $s_0 = ab_0d_2-cd_0=0$, and $s_1 = ab_1d_2-cd_1=0$.  Denote the corresponding generators of $\Ext^2$ by $\fd r_0$, $\fd s_0$, and $\fd s_1$.  We first pick maps of projective resolutions representing these generators, which all turn out to be uniquely determined.  We have the projective resolutions:

\begin{equation}
\xymatrix@R=3mm{  
&&0\ar[r]&P_0\ar[r]&L_0\ar[r]&0\\
&0\ar[r]&P_0\ar[r]&P_1\ar[r]&L_1\ar[r]&0\\
&0\ar[r]&P_0 \oplus P_1^{\oplus2}\ar[r]&P_2\ar[r]&L_2\ar[r]&0\\
0\ar[r]&P_0^{\oplus2} \oplus P_1\ar[r]&P_2^{\oplus3}\ar[r]&P_3\ar[r]&L_3\ar[r]&0
}
\end{equation}

We choose representatives of $\Ext^1(L_i,L_j)$ as follows:  for $i \leq 3$, there are no relations originating at the $i$'th node of the quiver and hence the maps representing $\fd b_i$, $\fd a$, and $\fd c$ are uniquely determined.  The choice of representative of $\fd d_i$ is uniquely determined as well.  That is to say, in the diagram

\begin{equation}
\xymatrix{
P_0^{\oplus2} \oplus P_1\ar[r]\ar[d]&P_2^{\oplus3}\ar[r]\ar[d]&P_3\\
P_0 \oplus P_1^{\oplus2}\ar[r]&P_2&
}
\end{equation}the bottom horizontal map is injective, so that the left vertical map is uniquely determined by the right vertical map, which we take to be projection on the $i$'th factor.  Finally, for the generators of $\Ext^2$ we take the obvious uniquely determined maps from the projective resolution of $L_3$ to the other $L_i$.  

We want to compute all products $m_k$ of the various $\Ext^1$'s.  Such products will all be in $\Ext^2$, and because only $\Ext^2(L_3,L_0)$ and $\Ext^2(L_3,L_1)$ are nonzero we see that the possible nonzero products are $m_2({\fd b_i}, {\fd d_j})$, $m_2({\fd c}, {\fd d_i})$, and $m_3({\fd a}, {\fd b_i}, {\fd d_j})$.  Let's look at the first of these; the relevant composition is:

\begin{equation}
\xymatrix{
&P_0^{\oplus2} \oplus P_1\ar[r]\ar[d]&P_2^{\oplus3}\ar[r]\ar[d]&P_3\\
&P_0 \oplus P_1^{\oplus2}\ar[r]\ar[d]&P_2\\
P_0\ar[r]&P_1}
\end{equation}

The map from the first to the second row is ${\fd d_j}$ and that from
the second to the third row is ${\fd b_i}$.  For convenience, we label
the individual $P_k$s that occur in various parts of the diagram.  The
upper left entry is $P_0^{\oplus2} \oplus P_1$, where each summand
corresponds to a different relation.  We naturally label the two
$P_0$s as $S_0$ and $S_1$, and we label the $P_1$ as $R_0$.  The entry
below this one is $P_0 \oplus P_1^{\oplus2}$, and here each $P_k$
corresponds to an arrow emanating from $P_2$.  Thus we label the $P_0$
as $C$ and the two $P_1$s as $B_0$ and $B_1$.

Let us consider the possible maps we can have.  For $k>l$ there are no
nonzero maps from $P_k$ to $P_l$.  From each $P_k$ to itself there is
the identity map, and it is the only one that will be of use to us.
For $k<l$, however, there are several ways to map $P_k$ to $P_l$, each
corresponding to a path from node $l$ to node $k$.  Thus, for example,
there is one map from $P_0$ to $P_1$, denoted by $a$.

To see how ${\fd d_j}$ acts note that it maps $R_0 \oplus S_0 \oplus
S_1$ to $C \oplus B_0 \oplus B_1$.  We can thus represent its action
on $R_0 \oplus S_0 \oplus S_1$ as a 3 by 3 matrix.  From the
definition of ${\fd d_j}$ it is easy to obtain (by slight abuse of
notation we denote by ${\fd d_j}$ both itself and its restriction to
$R_0 \oplus S_0 \oplus S_1$):

\begin{equation}
{\fd d_0} = \left(\begin{matrix}0&{-1}&0\\0&0&0\\{-1}&0&0\end{matrix}\right),
{\fd d_1} = \left(\begin{matrix}0&0&{-1}\\1&0&0\\0&0&0\end{matrix}\right),
{\fd d_2} = \left(\begin{matrix}0&0&0\\0&a&0\\0&0&a\end{matrix}\right).
\end{equation} 
Note that these matrix elements are simply obtained by ``contracting''
the relevant relation (which indexes the column) with $d_j$ to obtain
the elements of the column.  Similar reasoning shows that the ${\fd
b_i}$ act as follows:

\begin{equation}
{\fd b_0} = \left(\begin{matrix}0&1&0\end{matrix}\right),
{\fd b_1}= \left(\begin{matrix}0&0&1\end{matrix}\right).
\end{equation}The nonzero compositions are

\begin{equation}
{\fd b_0} {\fd d_1} = \left(\begin{matrix}1&0&0\end{matrix}\right),
{\fd b_0} {\fd d_2} = \left(\begin{matrix}0&a&0\end{matrix}\right),
\end{equation}

\begin{equation}
{\fd b_1} {\fd d_0} = \left(\begin{matrix}{-1}&0&0\end{matrix}\right),
{\fd b_1} {\fd d_2} = \left(\begin{matrix}0&0&a\end{matrix}\right),
\end{equation}Now, note that the compositions ${\fd b_0} {\fd d_2}$ and ${\fd b_1} {\fd d_2}$ can both be factored through the leftmost map $P_0 \rightarrow P_1$, and hence are exact in the quiver dga.  So the only nonzero products are

\begin{equation}
m_2 ({\fd b_0}, {\fd d_1}) = {\fd r_0},
m_2 ({\fd b_1}, {\fd d_0}) = -{\fd r_0}.
\end{equation}A good shorthand way of expressing this result is that $m_2 ({\fd b_i}, {\fd d_j}) = (b_i d_j, R_0) {\fd r_0} + (b_i d_j, S_0) {\fd s_0} + (b_i d_j, S_1) {\fd s_1}$, where the parentheses denote the coefficient of the string represented by the left argument in the relation represented by the right argument.  As one traces through the above manipulations it is clear that this is a general result that always holds when one computes the products $m_2$.  Thus we have $m_2 ({\fd c, \fd d_0}) = -{\fd s_0}$, $m_2 ({\fd c, \fd d_1}) = -{\fd s_1}$, and $m_2({\fd c, \fd d_2}) = 0$.

To compute $m_3$, we first have to define a choice of $f_2$, which must satisfy 
\begin{equation}
im_2 = (i\cdot i) + df_2.
\end{equation}Quick inspection shows that we may take $f_2 = 0$ everywhere except for $f_2 ({\fd b_i, \fd d_j})$.  From the above analysis, we see that 

\begin{equation}{
df_2({\fd b_0, \fd d_2}) = \left(\begin{matrix}0&{-a}&0\end{matrix}\right),
df_2({\fd b_1, \fd d_2}) = \left(\begin{matrix}0&0&{-a}\end{matrix}\right).
}
\end{equation}where $a$ denotes right multiplication and the notation indicates a map $P_0^{\oplus2} \oplus P_1 \rightarrow P_1$.  Hence we can define $f_2({\fd b_0, \fd d_2})$ and $f_2({\fd b_1, \fd d_2})$ respectively as:

\begin{equation}
\xymatrix{  
P_0^{\oplus2} \oplus P_1\ar[r]\ar[d]^-{\left(\begin{smallmatrix}0&-1&0\end{smallmatrix}\right)}&P_2^{\oplus3}\ar[r]\ar[d]^0&P_3\\
P_0\ar[r]&P_1
}
\end{equation} 
\begin{equation}
\xymatrix{  
P_0^{\oplus2} \oplus P_1\ar[r]\ar[d]^-{\left(\begin{smallmatrix}0&0&-1\end{smallmatrix}\right)}&P_2^{\oplus3}\ar[r]\ar[d]^0&P_3\\
P_0\ar[r]&P_1}
\end{equation}Now, we have 

\begin{equation}
im_3 = f_2(\id\otimes m_2)-f_2(m_2\otimes\id) + (i\cdot f_2) -(f_2\cdot i) + d f_3,
\end{equation}so that, recalling the sign rule (\ref{eq:sign_rule}), $m_3 ({\fd a, \fd b_i, \fd d_j}) = -\left[ {\fd a} \cdot f_2 ({\fd b_i, \fd d_j}) \right] $.  Composing with ${\fd a}$ we see that $m_3({\fd a, \fd b_0, \fd d_2}) = {\fd s_0}$ and $m_3({\fd a, \fd b_1, \fd d_2}) = {\fd s_1}$.  Again, a shorthand way of expressing this result is $m_3 ({\fd a, \fd b_i, \fd d_j}) = (a b_i d_j, R_0) {\fd r_0} + (a b_i d_j, S_0) {\fd s_0} + (a b_i d_j, S_1) {\fd s_1}$.  These are all the nonzero $A_\infty$ products in this example.  

One can see by carrying out the $A_\infty$ algorithm that this formula generalizes to all the $m_k$ in any quiver.  We sketch the argument modulo various signs, which have to be checked carefully.  Let the relations by labeled by $R_i$ and the corresponding generators of $\Ext^2$ by ${\fd r_i}$, as above.  One proceeds by induction on $k_0$.  Let's take the following inductive hypothesis:  for all $j<k_0$, we have

\begin{equation}
m_j(a_1, \ldots, a_j) = \sum_i (a_1 \ldots a_j, R_i) {\fd r_i} \label{eq:ind}
\end{equation}as well as a statement about $f_k$ for which we need to introduce some notation.  Each $a_i$ is an $\Ext^1$, so that it can be represented as a map between the projective resolution of $L_{m(i)}$ and $L_{m(i-1)}$.  According to this notation, $a_i$ is an arrow between node $m(i)$ and $m(i-1)$.  The projective resolution of $L_{m(i)}$ is

\begin{equation}
\xymatrix@1{
\ldots\ar[r]&\bigoplus_{k}P_k^{\oplus r_{m(i)k}}
  \ar[r]&\bigoplus_{k}P_k^{\oplus n_{m(i)k}}\ar[r]&P_{m(i)}\ar[r]&L_{m(i)}\ar[r]&0.
}
\end{equation}
The second part of the inductive hypothesis is that for $j<k_0$,
$f_j(a_1, \ldots, a_j)$ is represented as:

\begin{equation}
\xymatrix{  
\bigoplus_{k}P_k^{\oplus r_{m(j)k}}\ar[r]\ar[d]&
  \bigoplus_{k}P_k^{\oplus n_{m(j)k}}\ar[r]\ar[d]^0&P_{m(j)}\\
\bigoplus_{k}P_k^{\oplus n_{m(0)k}}\ar[r]&P_{m(0)}} \label{eq:f_j}
\end{equation}
where the left vertical map takes each relation to an $\Ext^1$
determined by the contraction of the relation with $a_1 \ldots a_j$
(we extract only the linear terms in the contraction, as only these
correspond to $\Ext^1$s).  $f_j$ with any $\Ext^2$s as arguments
vanish.

For the inductive step, we have to prove the analogous statements for
$m_{k_0}$.  We use (\ref{eq:Amorph}) to write $m_{k_0}$ in terms of
the lower $m_j$s and $f_j$s.  We then note from the above form of the
$f_j$ that all terms vanish except $(i\cdot f_{k_0-1})$ --- basically
because the only nonzero map in (\ref{eq:f_j}) takes relations to
$\Ext^1$s, so the composition of two $f_j$s vanishes.  This
straightforwardly leads to (\ref{eq:ind}) for $m_{k_0}$.  Inverting
$d$ shows that $f_{k_0}$ may be chosen to be of the form
(\ref{eq:f_j}).

Thus essentially one knows these products as soon as one knows all the
relations in the quiver.  It follows that the term in the
superpotential that multiplies the ``extra'' $\Ext^1$ corresponding to
a given relation is simply that relation (written as a polynomial in
the $\Ext^1$s).  We can now write down the superpotential:

\begin{equation}
W = R_0 (B_0 D_1 - B_1 D_0) + S_0 (A B_0 D_2 - C D_0) + S_1 (A B_1 D_2 - C D_1)
\end{equation}

\section{Conclusions}

We have given an effective method for computing superpotentials for
quiver gauge theories associated with shrinking del Pezzo cycles.  We
showed that the superpotential is linear in the fields that correspond
to $\Ext^2$s in the del Pezzo quiver, and that each such field
multiplies a polynomial which is just the corresponding relation.  To
do this we performed a precise reduction of the problem from one
involving sheaves on the Calabi-Yau to one involving sheaves on the
del Pezzo.  We solved the problem on the del Pezzo by switching to the
algebraically more tractable category of quiver representations and
explicitly evaluating the $A_\infty$ products there.  We did only the
cases where $S$ is $\P^2$ and $\dP 1$, but these examples show that
the algorithm is trivial to carry out provided one has the quiver and
the relations.  These, of course, might not be so trivial to obtain,
especially for the higher del Pezzos, which themselves have complex
structure moduli.  These complex structure moduli are contained in the
choice of points to be blown up on $\P^2$, and will show up in the
quiver relations.

%%%%%%%%%%%%%%%%%%%%%%%%%%%%%%%%%%%%%%%%%%%%%%%%%%%%%%%%%%%%%%%%%%%

\section*{Acknowledgments}

We wish to thank X.~Liu, J.~McGreevy, A.~Saltman, A.~Tomasiello for
useful conversations. P.S.A.~is supported in part by NSF grant
DMS-0301476, Stanford University, SLAC and the Packard Foundation.
L.M.K.~is supported by Stanford University.

%\bibliographystyle{my-phys}
%\bibliography{string}

\end{document}